\documentclass[twocolumn]{webofc}

\usepackage[varg]{txfonts}   
\usepackage{float}   
\begin{document}
\title{Neutrinoless Double-Beta Decay and Realistic Shell Model}
%
%

\author{\firstname{Nunzio} \lastname{Itaco}\inst{1,2}\fnsep\thanks{\email{nunzio.itaco@unicampania.it}} \and
        \firstname{Luigi} \lastname{Coraggio}\inst{2} \and
        \firstname{Riccardo} \lastname{Mancino}\inst{1,2}
}

\institute{
Dipartimento di Matematica e Fisica, Universit\`a degli
  Studi della Campania ``Luigi Vanvitelli'', viale Abramo Lincoln 5 -
  I-81100 Caserta, Italy
\and
Istituto Nazionale di Fisica Nucleare, \\
Complesso Universitario di Monte  S. Angelo, Via Cintia - I-80126 Napoli, Italy   
          }

\abstract{
We report on the calculation of the neutrinoless
double-$\beta$ decay nuclear matrix element for $^{76}$Ge within the framework of the
realistic shell model. The effective shell-model Hamiltonian and the
two-body transition operator describing the decay are derived by way
of many-body perturbation theory.
Particular attention is focused on the role played by the so-called Pauli
blocking effect in the derivation of the effective operator.
}
\maketitle
\section{Introduction}
\label{intro}
Neutrinoless double-$\beta$ ($0\nu\beta\beta$) decay is an exotic nuclear
process predicted by extensions of the Standard Model of particle
physics. Observation of such a process would prove
the non-conservation of lepton number, and show that neutrinos
have a Majorana mass component (see Refs. \cite{Avignone08,Vergados12} and references
therein).

In the framework of light Majorana neutrino exchange, the half life of the
$0\nu\beta\beta$ decay is inversely proportional to the square of the
effective Majorana neutrino mass $\langle m _{\nu} \rangle \equiv |
\sum_i U_{ei}^2 m_{\nu_i}|$.
Explicitly, it can be written as 
\begin{equation}
\left[ T^{0\nu}_{1/2} \right]^{-1} = G^{0\nu} \left| M^{0\nu}
\right|^2 \langle m _{\nu} \rangle^2 ~~,
\label{halflife}
\end{equation}
\noindent
where $G^{0\nu}$ is the so-called phase-space factor (or kinematic
factor), and $M^{0\nu}$ is the nuclear matrix element (NME) directly related to the
wave functions of the parent and grand-daughter nuclei.
At present, the phase-space factors involved in the double $\beta$
decays of nuclei of experimental interest are calculated with great
accuracy \cite{Kotila12,Stoica13}, therefore it is crucial to have
precise values of the NME, both to improve the reliability of the
$0\nu\beta\beta$ lifetime predictions, and to extract neutrino
properties from the experimental results.

Several nuclear structure models have been employed to provide
NME values as precise as possible, the most largely employed being, at present,  
the Interacting Boson Model (IBM)
\cite{Barea09,Barea12,Barea13}, the Quasiparticle Random-Phase
Approximation (QRPA) \cite{Simkovic08,Simkovic09,Fang11,Faessler12},
Energy Density Functional methods \cite{Rodriguez10}, and the Shell
Model (SM)
\cite{Caurier08,Menendez09a,Menendez09b,Horoi13a,Horoi13b,Neacsu15,Brown15}. 

Each model can be more suitable than another for a certain class of
nuclei, and, when comparing the calculated NMEs, it can be
seen that, at present, the results obtained with different approaches
can differ by a factor of two or three (see for instance the review \cite{Engel17}). 

It is worth noting that, all the above models are based on the
use of a truncated Hilbert space to overcome the computational
complexity, but only within the realistic shell model an effective
$0\nu\beta\beta$ operator, which takes into account the degrees of
freedom which do not appear explicitly in the calculated
wavefunctions, may be introduced.
This approach has been pioneered by Kuo and coworkers
\cite{Song89,Staudt92}, and more recently pursued by Holt and Engel
\cite{Holt13}.
Actually, in the SM approach it is possible to consider the
correlations involving single-particle orbitals outside the model
space deriving an effective operator.
This is the so-called realistic shell-model approach where both the
effective Hamiltonian and the transition operators are derived from a
realistic free nucleon-nucleon ($NN$) potential $V_{NN}$ by way of the many-body theory
\cite{Kuo90,Suzuki95}.

In two recent papers \cite{Coraggio17a,Coraggio19}, we have adopted the
above approach to calculate
properties related to the GT and $2\nu\beta\beta$ decays of
nuclei $0\nu\beta\beta$-candidates, with mass ranging from
$A=48$ to $A=136$. Our results and their comparison with the
experimental data have shown that many-body perturbation theory (MBPT)
allows to derive effective SM Hamiltonians and transition operators
that reproduce quantitatively the observed spectroscopic and decay 
properties, without resorting to an empirical quenching of the axial
coupling constant $g_A$.

On these grounds, we report in this paper on a preliminary calculation of the
NME involved in the $0\nu\beta\beta$ decay of $^{76}$Ge, that is
currently investigated by GERDA experiment \cite{GERDA} at the Laboratori Nazionali
del Gran Sasso (LNGS) of INFN, and by CDEX-1 experiment \cite{CDEX} at China
Jinping Underground Laboratory.   

Our theoretical framework is the many-body perturbation theory
\cite{Kuo81,Suzuki95,Coraggio12a,Coraggio17a}; we start from a realistic
nuclear potential, and derive an effective SM Hamiltonian and
a $0\nu\beta\beta$ decay operator that are employed to calculate the
wavefunctions of the ground states of $^{76}$Ge and $^{76}$Se and the
$0\nu\beta\beta$ NME.
We focus our attention on the role of the so-called ``blocking
effect'', that takes into account the Pauli 
exclusion principle in systems with more than two valence nucleons
\cite{Ellis77,Towner87} in the derivation of the two-body effective
$0\nu\beta\beta$ operator.

The paper is organized as follows. In the next section we give a brief
description of how we derive the shell-model Hamiltonian and the two-body effective
$0\nu\beta\beta$ decay operator.
In Section \ref{results}, we report the results that we obtain for the $^{76}$Ge
NME, analyzing in detail the contribution arising from the core-polarization
blocking diagrams. A brief summary is reported in the last section.

\section{Outline of calculations}
\label{outline}
The starting point of our calculation is provided by the high-precision CD-Bonn
$NN$ potential \cite{Machleidt01b}, that is smoothed integrating out
its repulsive high-momentum components by
way of the so-called $V_{\rm low-k}$ approach \cite{Bogner01,Bogner02}.
In this way we get a softer
$NN$ potential defined up to a cutoff $\Lambda$, that preserves the
physics of the original CD-Bonn interaction.
The value of $\Lambda$ is chosen, as in many of our recent papers
\cite{Coraggio17a,Coraggio19,Coraggio15a,Coraggio15b,Coraggio16a}
equal to $2.6$ fm$^{-1}$, this value being a trade off between the
need of minimizing the role of the missing three-nucleon force (3NF)
\cite{Coraggio15b} and that of ensuring the perturbative behavior of the potential. 
The Coulomb potential is explicitly taken into account in the
proton-proton channel.

The harmonic oscillator (HO) potential $U$ is introduced as
an auxiliary one-body potential in order to break up the Hamiltonian
for a system of $A$ nucleons as the
sum of a one-body term $H_0$, which describes the independent motion
of the nucleons, and a residual interaction $H_1$:

\begin{eqnarray}
 H &= & \sum_{i=1}^{A} \frac{p_i^2}{2m} + \sum_{i<j=1}^{A} V_{\rm low-k}^{ij}
 = T + V_{\rm low-k} = \nonumber \\
~& = & (T+U)+(V_{\rm low-k}-U)= H_{0}+H_{1}~~.\label{smham}
\end{eqnarray}

\noindent
It is now possible to define a truncated model space in terms of the
eigenvectors of $H_0$. To study the $0\nu\beta\beta$ decay of
$^{76}$Ge we employ a model space spanned by the four 
$1p_{3/2},1p_{1/2},0f_{5/2},0g_{9/2}$
proton and neutron orbitals outside the doubly-closed $^{56}$Ni
core, and we derive an effective shell-model
Hamiltonian $H_{\rm eff}$, that takes into account the degrees
of freedom that are not explicitly included in the shell-model
framework.

We derive $H_{\rm eff}$ by resorting to the many-body perturbation
theory, an approach that has been developed by Kuo and coworkers
through the 1970s \cite{Kuo90,Kuo95}.
More precisely, we use the well-known $\hat{Q}$
box-plus-folded-diagram method \cite{Kuo71}, where the $\hat{Q}$ box
is defined as a function of the unperturbed energy $\epsilon$ of the valence particles:
\begin{equation}
\hat{Q}(\epsilon) = P H_1 P + P H_1 Q \frac{1}{\epsilon - QHQ} Q H_1 P~~,
\label{qbox}
\end{equation}

\noindent
where the operator $P$ projects onto the model space and $Q=\mathbf{1}
-P$.
In the present calculations the $\hat{Q}$ box is expanded as a
collection of one- and two-body irreducible valence-linked Goldstone
diagrams up to third order in the perturbative
expansion\cite{Coraggio10a,Coraggio12a}.

Within this framework, it can be shown that the effective Hamiltonian $H_{\rm eff}$ can be
written in terms of the $\hat{Q}$ box derivatives \cite{Krenciglowa74}

\begin{equation}
\hat{Q}_m = \frac {1}{m!} \frac {d^m \hat{Q} (\epsilon)}{d \epsilon^m} \biggl| 
_{\epsilon=\epsilon_0} ~~, 
\label{qm}
\end{equation}

\noindent

\begin{equation}
H_{\rm eff} = \sum_{i=0}^{\infty} F_i~~,
\label{kkeq}
\end{equation}

\noindent
where

\begin{eqnarray}
F_0 & = & \hat{Q}(\epsilon_0)  \nonumber \\
F_1 & = & \hat{Q}_1(\epsilon_0)\hat{Q}(\epsilon_0)  \nonumber \\
F_2 & = & \left[ \hat{Q}_2(\epsilon_0)\hat{Q}(\epsilon_0) + 
\hat{Q}_1(\epsilon_0)\hat{Q}_1(\epsilon_0) \right] \hat{Q}(\epsilon_0)  \nonumber \\
~~ & ... & ~~ 
\label{kkeqexp}
\end{eqnarray}

\noindent
$\epsilon_0$ being the model-space eigenvalue of the unperturbed
Hamiltonian $H_0$.

The $H_{\rm eff}$ for one-valence-nucleon systems provides the
single-particle (SP) energies for our SM calculations, while the
two-body matrix elements (TBMEs) are obtained from $H_{\rm eff}$
derived for the nuclei with two valence nucleons, by subtracting the
theoretical SP energies.

The so obtained SP energies and TBMEs can be found in
\cite{Coraggio19}, where it is also reported a detailed discussion of the
perturbative properties of $H_{\rm eff}$. 

The NME involved in $0\nu\beta\beta$ decay, $M^{0\nu}$ , may be
expressed as
\begin{equation}
 M^{0 \nu}=M^{0 \nu}_{GT}-\left( \frac{g_V}{g_A} \right)^2  M^{0
   \nu}_F 
\end{equation}
\noindent
where the matrix elements are defined as follows:
\begin{equation}
M_\alpha^{0\nu} = \sum_{m,n} \left< 0^+_f \mid \tau^-_{m} \tau^-_{n} O^\alpha_{mn}\mid 0^+_i \right> \ ,
\end{equation}
with $\alpha=(GT,\ F)$, having neglected the tensor term, whose effect
has been estimated to be of the order of few percent \cite{Menendez09b}. 
The operators $O^\alpha_{mn}$ have the following expression:
\begin{eqnarray}
\nonumber O_{mn}^{GT} & = & \vec{\sigma}_m \cdot \vec{\sigma}_n H_{GT}(r) \ , \\
\nonumber O_{mn}^{F} & = & H_{F}(r)\ ,  
\end{eqnarray}
and the neutrino potentials $H_\alpha$ are defined using the closure approximation
\begin{equation}
H_\alpha(r) =  \frac{2R}{\pi}\int^\infty_0 f_{\alpha}(qr)
\frac{h_\alpha(q^2)}{q+\left< E \right> }G_\alpha(q^2) q dq  ,
\end{equation}
\noindent
where $f_{F,GT}(qr)=j_0(qr)$, 
$\left< E \right>$ is the average energy of the virtual intermediate
states used in the closure approximation, while the explicit
expression of the form factors $h_\alpha(q^2)$ can be found for
instance in Ref. \cite{Senkov13}.

As mentioned before, we want to derive the effective $0\nu\beta\beta$
decay operator tailored for the chosen model space. To this end, we
resort to the formalism presented by Suzuki and Okamoto in
Ref. \cite{Suzuki95}.
In this approach, a non-Hermitian effective operator $\Theta_{\rm
  eff}$ can be expressed as 
\begin{eqnarray}
\Theta_{\rm eff} & = & (P + \hat{Q}_1 + \hat{Q}_1 \hat{Q}_1 + \hat{Q}_2
\hat{Q} + \hat{Q} \hat{Q}_2 + \cdots)\nonumber \\
~ & ~& \times (\chi_0+ \chi_1 + \chi_2 +\cdots)~~. \label{effopexp1}
\end{eqnarray}
where the operators $\chi_n$ are defined as:
\begin{eqnarray}
\chi_0 &=& (\hat{\Theta}_0 + h.c.)+ \hat{\Theta}_{00}~~,  \label{chi0} \\
\chi_1 &=& (\hat{\Theta}_1\hat{Q} + h.c.) + (\hat{\Theta}_{01}\hat{Q}
+ h.c.) ~~, \\
\chi_2 &=& (\hat{\Theta}_1\hat{Q}_1 \hat{Q}+ h.c.) +
(\hat{\Theta}_{2}\hat{Q}\hat{Q} + h.c.) + \nonumber \\
~ & ~ & (\hat{\Theta}_{02}\hat{Q}\hat{Q} + h.c.)+  \hat{Q}
\hat{\Theta}_{11} \hat{Q}~~, \label{chin} \\
&~~~& \cdots \nonumber
\end{eqnarray}

\noindent
where $\hat{\Theta}_m$, $\hat{\Theta}_{mn}$ have the following
expressions:
\begin{eqnarray}
\hat{\Theta}_m & = & \frac {1}{m!} \frac {d^m \hat{\Theta}
 (\epsilon)}{d \epsilon^m} \biggl|_{\epsilon=\epsilon_0} ~~~, \\
\hat{\Theta}_{mn} & = & \frac {1}{m! n!} \frac{d^m}{d \epsilon_1^m}
\frac{d^n}{d \epsilon_2^n}  \hat{\Theta} (\epsilon_1 ;\epsilon_2)
\biggl|_{\epsilon_1= \epsilon_0, \epsilon_2  = \epsilon_0} ~,
\end{eqnarray}

\noindent
with
\begin{eqnarray}
\hat{\Theta} (\epsilon) = & P \Theta P + P \Theta Q
\frac{1}{\epsilon - Q H Q} Q H_1 P ~, ~~~~~~~~~~~~~~~~~~~\label{thetabox} \\
\hat{\Theta} (\epsilon_1 ; \epsilon_2) = & P H_1 Q
\frac{1}{\epsilon_1 - Q H Q} Q \Theta Q \frac{1}{\epsilon_2 - Q H Q} Q H_1 P ~,~~~~~
\end{eqnarray}

\noindent
$\Theta$ being the bare transition operator.

It is worth noting that using Eqs. \ref{kkeq}, \ref{kkeqexp},
Eq. \ref{effopexp1} may be then rewritten as
\begin{equation}
\Theta_{\rm eff}  = H_{\rm eff} \hat{Q}^{-1}  (\chi_0+ \chi_1 + \chi_2 +\cdots) ~~,
\label{effopexp2}
\end{equation}
enlightening the connection existing between the effective Hamiltonian
and the effective operators. 

\begin{figure}[h]
\centering
\includegraphics[scale=0.23]{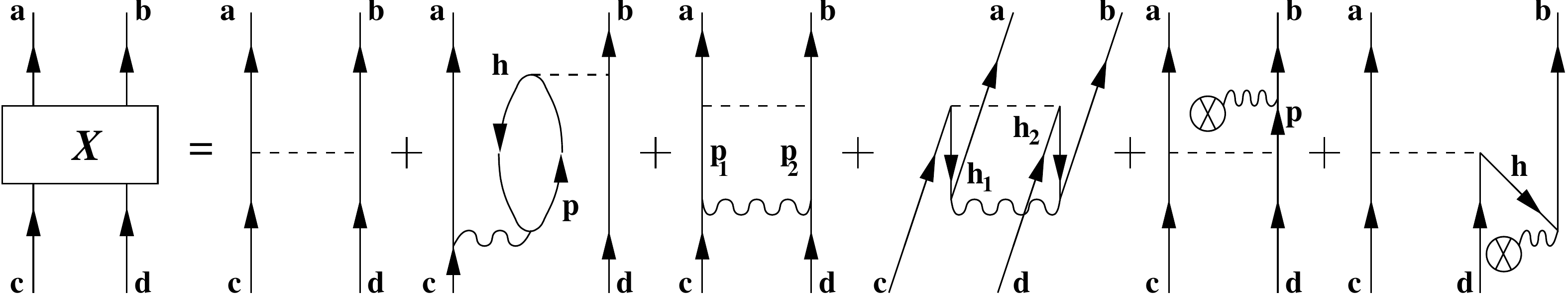}
\caption{Two-body second-order diagrams included in the perturbative
  expansion of $\hat{\Theta}$. The dashed lines indicate the bare operator
  $\Theta$, the wavy lines the two-body potential $V_{\rm low-k}$.}
\label{figeffop2}  
\end{figure}

In present calculation for the effective two-body $0\nu\beta\beta$
decay operator, we arrest the $\chi_n$ series to the $\chi_2$ term. 
Since $\chi_3$ depends on the first, second, and third derivatives of
$\hat{\Theta}_0$ and $\hat{\Theta}_{00}$, and on the first and second
derivatives of the $\hat{Q}$ box (see Eq. \ref{chin}), our estimation
of these quantities leads to evaluate $\chi_3$ being at least one
order of magnitude smaller than $\chi_2$.
The calculation is performed starting from a perturbative expansion of
$\hat{\Theta}_0$ and $\hat{\Theta}_{00}$, and in Fig. \ref{figeffop2}
we report all the two-body $\Theta_0$ diagrams up to the first order
in $V_{\rm low-k}$, the bare operator $\Theta$ being represented by a dashed line.

The circle with a cross inside represents a first-order 
$(V_{\rm  low-k}-U)$-insertion, arising from the $-U$ 
term in the interaction Hamiltonian $H_1$ (see for example
Ref. \cite{Coraggio12a} for details).

\begin{figure}[h]
\begin{center}
\includegraphics[scale=0.60,angle=0]{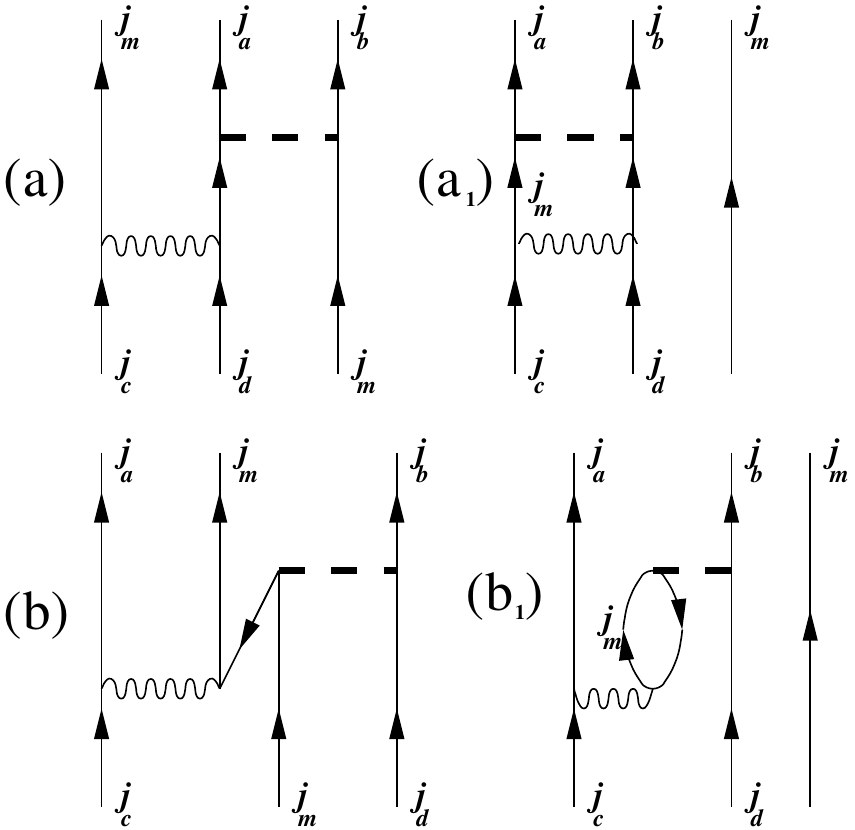}
\caption{Three-body second-order diagrams that should appear in the
  perturbative expansion of $\hat{\Theta}$. As in
  Fig. \ref{figeffop2}, the dashed line indicates the bare operator 
  $\Theta$, the wavy line the two-body potential $V_{\rm low-k}$. For
  the sake of simplicity, for each topology only one of the 
diagrams which correspond to the permutation of the external lines is drawn. }
\label{figeffop3}
\end{center}
\end{figure}

The diagrams in Fig. \ref{figeffop2} refer to the derivation of the
effective operator for a system with two valence-nucleons. When
dealing with nuclei with a larger number of valence nucleons,
many-body diagrams come into play, accounting for the interaction via
the two-body force of the many-valence nucleons with core excitations
as well as with virtual intermediate nucleons scattered above the
model space.
The two topologies of second-order connected three-valence-nucleon
diagrams  for a two-body operator $\hat{\Theta}$ are reported in
Fig. \ref{figeffop3} (diagrams (a) and (b)).
These diagrams correct the  Pauli-principle violation introduced by
diagram ($\rm{a}_1$) and ($\rm{b}_1$) when one of the intermediate particle
states is equal to $m$.

Since there are no shell-model codes, at present, able to handle
three-body transition operators, we derive a density-dependent 
two-body operator from the three-body one by summing and averaging over
one incoming and outcoming particle of the connected diagrams
($\mathrm{a}$) and ($\mathrm{b}$) of Fig. \ref{figeffop3}.
In this way we take into account the filling of the model-space
orbitals when dealing with more than two valence nucleons.

More precisely, first we calculate exactly the three-body connected
diagrams ($\mathrm{k}$) 
$\langle [ (j_a j_b)_J , j_m]_{J'}| (k) |[ (j_c j_d)_J ,
j_n]_{J'}\rangle$, with $\mathrm{k}$  equal to $\mathrm{a}$ and
$\mathrm{b}$, and whose expressions can be found in
Ref. \cite{Polls83}, then we evaluate the corresponding
density-dependent two-body diagram $\langle [ (j_a j_b)_J |
(\overline{k}) |[ (j_c j_d)_J \rangle$, whose explicit expression is

\begin{eqnarray}
\label{diagramdd}
\langle [ (j_a j_b)_J | (\overline{k}) |[ (j_c j_d)_J \rangle = ~~~~~~~~~~~~~~~~~~~~~~~~~~~~~~~~~~~
  \nonumber \\
\sum_{m,J'} \rho_m \frac{\hat{J}'^2}{\hat{J}^2} \langle [ (j_a j_b)_J , j_m]_{J'}| (k) |[ (j_c j_d)_J , j_m]_{J'}\rangle
\end{eqnarray}
where the summation over $m$-index runs in the model
space, and $\rho_m$ is the unperturbed occupation density of the orbital $j_m$
accordingly to the number of valence nucleons.

The so obtained two-body contribution is finally added to the
collection of the diagrams of the perturbative expansion of
$\hat{\Theta}$. 

\section{Results}
\label{results}
In this Section we present the results of our SM calculations for the
NME of $^{76}$Ge.
As previously mentioned, all the calculations have been performed employing theoretical SP
energies, TBMEs, and effective transition operators, describing
$^{76}$Ge as a core of $^{56}$Ni plus 4 protons and 16 neutrons
interacting in a reduced model space spanned by the  four
$1p_{3/2},1p_{1/2},0f_{5/2},0g_{9/2}$ orbitals.

In Ref. \cite{Coraggio19} we have compared the calculated low-energy
spectrum, the electromagnetic properties,  the GT$^-$ strength distributions and the
calculated NMEs of the $2\nu\beta\beta$ decay with the available
experimental data, showing a quantitative agreement.

In order to see how much the choice of the effective Hamiltonian can
affect the value of the NME, in Table \ref{NMEcomp} we compare
our calculated NME using the bare $0\nu\beta\beta$ operator without any
renormalization, with the values obtained in two other SM
calculations performed using the same model space but with different
empirical SM hamiltonians \cite{Menendez09b,Senkov14}.

\begin{table}[h]
\centering
\caption{Theoretical $0\nu\beta\beta$ NME calculated in the framework
  of the SM using different effective Hamiltonians. $I$ and $II$ are
  the results obtained by Menendez and coworkers \cite{Menendez09b}
  and by Sen'kov and coworkers
  \cite{Senkov14}, respectively. The $0\nu\beta\beta$ operator employed is not
renormalized.}
\label{NMEcomp}
\begin{tabular}{cccc}
\hline
& Our & I & II  \\
\hline
NME & 3.40 & 2.96 & 3.25 \\
\hline
\end{tabular}
\end{table}

As it can be seen, the theoretical results are in good agreement, the
largest discrepancy non exceeding 15$\%$.

In Table \ref{NMEresults} we report the calculated $0\nu\beta\beta$
NME obtained using an effective operator derived by second order MBPT and taking
into account all the intermediate states up to an excitation energy of
14 $\hbar \omega$, which are enough to provide convergent NME values.
To clarify the role of Pauli blocking diagrams we have
derived two different operators with and without taking into account
diagrams (a) and (b) of Fig. \ref{figeffop3}.
 
\begin{table}[h]
\centering
\caption{$M^{0 \nu}$, $M^{0 \nu}_{GT}$ and $M^{0 \nu}_{F}$ for
  $^{76}$Ge calculated using the $0\nu\beta\beta$ effective
  operator. See text for details.}
\label{NMEresults}
\begin{tabular}{ccc}
\hline
& 2nd & 2nd + Pauli blocking   \\
\hline
$M^{0 \nu}_{GT}$ & 2.39 & 2.03   \\
& & \\
$M^{0 \nu}_{F}$ &  -0.64 &  -0.66 \\
& & \\
$M^{0 \nu}$ &  2.79 & 2.44   \\
\hline
\end{tabular}
\end{table}

The inspection of Table \ref{NMEresults} shows that including Pauli
blocking diagrams provides a further reduction of $M^{0 \nu}$ by an
amount around 10$\%$, and that this effect is mainly due to the
reduction of the Gamow-Teller contribution.

To better understand the action of Pauli blocking diagrams, we report
in Table \ref{blocking} the values of the diagrams reported in
Fig. \ref{figeffop3} for GT-operator corresponding to the most relevant configurations
involved in the $^{76}$Ge $0\nu\beta\beta$ decay.

\begin{table}[h]
\centering
\caption{Values of the second-order diagrams reported in
  Fig. \ref{figeffop3}. The values of the diagrams (a) and (b) are
calculated according to the expression in Eq. \ref{diagramdd}}
\label{blocking}
\begin{tabular}{lcccc}
\hline
$(j_a,j_b)_{J=0}$ $(j_c,j_d)_{J=0}$ & ($\rm{a}$) & ($\rm{a}_1$) & ($\rm{b}$) & ($\rm{b}_1$)  \\
\hline
$(f_{5/2},f_{5/2})$ $(g_{9/2},g_{9/2})$ & 0.157 &
-0.337 & -1.096 & 0.335 \\
 & & & & \\
$(p_{3/2},p_{3/2})$ $(g_{9/2},g_{9/2})$ & 0.189 &
-0.263 & -0.219 & 0.058 \\
\hline
\end{tabular}
\end{table}

As expected, the density dependent diagrams (a) and (b) reduce the
contribution of the Pauli violating diagrams ($\rm{a}_1$) and ($\rm{b}_1$).

\section{Summary}
\label{summary}
In this paper we have reported on some preliminary results for the SM calculation of the
nuclear matrix element involved in the $^{76}$Ge $0\nu\beta\beta$ decay.  
The effective shell-model Hamiltonian and the
two-body transition operator describing the decay are derived by way
of many-body perturbation theory at third and second order, respectively.
The role of the Pauli blocking three body diagrams that
appear in the perturbative expansion of the effective operator when
dealing with more than two valence nucleons, is taken into account in
an approximate way by introducing two-body density-dependent diagrams.


\end{document}